# Migrating Thermal Tides in the Martian Atmosphere during Aphelion Season Observed by EMM/EMIRS

**Siteng Fan[1], François Forget[1], Michael D. Smith[2], Sandrine Guerlet[1], Khalid M. Badri[3], Samuel A. Atwood[4,5], Roland M. B. Young[6], Christopher S. Edwards[7], Philip R. Christensen[8], Justin Deighan[5], Hessa R. Al Matroushi[3], Antoine Bierjon[1], Jiandong Liu[1], Ehouarn Millour[1]**

[1]LMD/IPSL, Sorbonne Université, PSL Research Université, École Normale Supérieure, École Polytechnique, CNRS, Paris, France

[2]NASA Goddard Space Flight Center, Greenbelt, MD, USA

[3]Mohammed Bin Rashid Space Centre, Dubai, UAE

[4]Space and Planetary Science Center, and Department of Earth Sciences, Khalifa University, Abu Dhabi, UAE

[5]Laboratory for Atmospheric and Space Physics, University of Colorado Boulder, Boulder, CO, USA

[6]Department of Physics & National Space Science and Technology Center, United Arab Emirates University, Al Ain, UAE

[7]Department of Astronomy and Planetary Science, Northern Arizona University, Flagstaff, AZ, USA

[8]School of Earth and Space Exploration, Arizona State University, Tempe, AZ, USA

Corresponding author: Siteng Fan (sfan@lmd.ipsl.fr)

**Key Points:**

- Migrating thermal tides in the Martian atmosphere are analyzed using the first set of EMM/EMIRS observations.

- The observed amplitudes of diurnal and semi-diurnal tides agree well with GCM predictions, but with different phases and wavelengths.

- Ter-diurnal tide is detected with an amplitude of ~0.5K, due to the observations covering the full range of local times.

## Abstract

Temperature profiles retrieved using the first set of data of the Emirates Mars InfraRed Spectrometer (EMIRS) obtained during the science phase of the Emirates Mars Mission (EMM) are used for the analysis of migrating thermal tides in the Martian atmosphere. The selected data cover a solar longitude ($L_S$) range of 60°-90° of Martian Year (MY) 36. The novel orbit design of the Hope Probe leads to a good geographic and local time coverage that significantly improves the analysis. Wave mode decomposition suggests dominant diurnal tide and important semi-diurnal tide with maximal amplitudes of 6K and 2K, respectively, as well as the existence of ~0.5K ter-diurnal tide. The results agree well with predictions by the Mars Planetary Climate Model (PCM), but the observed diurnal tide has an earlier phase (3h), and the semi-diurnal tide has an unexpectedly large wavelength (~200km).

## Plain Language Summary

As a result of its small thickness, the Martian atmosphere experiences large temperature variations within each Martian day due to the incoming sunlight. Such rapid and large temperature variations excite waves propagating in the Martian atmosphere that highly influence winds, cloud formation, and dust transport. In this work, we use the atmospheric temperature measurements derived using observations obtained by an infrared spectrometer onboard the Hope Probe to analyze the diurnal temperature variations and the excited waves. The novel design of the spacecraft's orbit provides good data coverage in location and time, leading to the success of detailed analyses of the waves that propagate in the Martian atmosphere synchronously with the movement of the Sun, among which a new wave mode with a period of one third of a Martian day is detected. We compare the results with predictions provided by numerical simulations, and they show good agreements in the wave strengths, but the observed waves have different wavelengths and phases.

## 1 Introduction

Thermal tides play important roles in the Martian atmosphere. As a result of the low atmospheric heat capacity and the fast planetary rotation, the Martian atmosphere experiences large and rapid daily temperature variations. Primarily driven by solar insolation and influenced by topography, thermal tides are excited as forms of planetary-scale harmonic responses (Gierasch & Goody 1968, Zurek 1976). Some modes of the tides propagate vertically with increasing amplitude due to the conservation of energy (Lindzen & Chapman 1969). These tides are significant and highly coupled with airborne dust and water ice as well as atmospheric circulation, and are sensitive indicators of excitation sources (Barnes et al. 2017, Wu et al. 2022). These excitations and coupling processes are among current bottlenecks of numerical simulations (Navarro et al. 2017, Gilli et al. 2020), and also our understanding of the Martian atmosphere.

Since the first global and diurnal observation of the Martian atmospheric temperature obtained by the Infrared Thermal Mapper (IRTM) onboard the Viking orbiters (Wilson & Richardson 2000), our knowledge of thermal tides in the Martian atmosphere has been significantly enriched in the past two decades from a number of Mars orbiter and lander observations (Hess et al., 1977, Banfield et al. 2003, Lee et al. 2009, Kleinböhl et al. 2013, Forbes et al. 2020). However, the local time coverage of most observations prevents detailed analysis of such planetary-scale diurnal/sub-diurnal variations. Sun-synchronous spacecraft orbits limit observations near two local times that result in wave mode aliasing, e.g., the Thermal Emission Spectrometer onboard the Mars Global Surveyor (MGS/TES, Banfield et al. 2003), and the Mars

Climate Sounder onboard the Mars Reconnaissance Orbiter (MRO/MCS, Lee et al. 2009, Kleinböhl et al. 2013, Forbes et al. 2020), while slow-drifting orbits introduce seasonal changes into diurnal variation analyses, e.g., the Planetary Fourier Spectrometer onboard the Mars Express (MEX/FPS, Giuranna et al. 2021), and the TIRVIM Fourier-spectrometer, part of the Atmospheric Chemistry Suite onboard the ExoMars Trace Gas Orbiter (TGO/ACS/TIRVIM, Fan et al. 2022). Therefore, data with planetary-scale spatial coverage that sample all local times within a short range of season is necessary for detailed thermal tide investigations. Observations obtained by the Emirates Mars InfraRed Spectrometer onboard the Hope Probe of the Emirates Mars Mission (EMIRS/EMM, Almatroushi et al. 2021, Edwards et al. 2021) meet such a requirement, which is the subject of this work.

## 2 Observations and Data Processing

### 2.1 EMM/EMIRS

The scientific objectives of EMM mainly focus on the Martian atmosphere (Almatroushi et al. 2021). The Hope Probe is in a high orbit (19970/42650km altitude at periapsis/apoapsis) with a low inclination (25°), which allows it to have a global-scale view of Mars from any location of the orbit. EMIRS is a Fourier transform infrared spectrometer onboard the spacecraft, which covers a spectral range of 1666-100cm$^{-1}$ with a resolution of 5 or 10cm$^{-1}$ depending on observing modes (Edwards et al. 2021). The instrument is equipped with a moving pointing mirror that samples the Martian disk within 0.5h, and a full geographic and local time coverage within 10 days (Figure 1a). Retrievals using these spectra provide information about the surface and atmospheric temperatures, water vapor, and dust and water ice aerosols (Smith et al. 2022), which use a constrained linear inversion method based on that for TES (Conrath et al. 2000, Smith et al. 2001, Smith 2002 and 2004) to fit sequentially for atmospheric temperature, aerosol optical depth and surface temperature, and water vapor, with first guesses of surface and atmospheric temperatures from the spectra themselves. The atmospheric temperature profiles are constrained from the Martian surface to ~50km (~2Pa) with a vertical resolution of approximately one scale height (~10km), and uncertainties ranging from ~2K at 1-3 scale heights above the surface to 5-10K at lower and higher altitudes (Smith et al. 2022).

### 2.2 Data Processing

Observations used in this work are taken from the first set of data obtained by EMIRS. These data range from the start of the EMM science phase in May 2021 to the Mars solar conjunction in September 2021, equivalent to a solar longitude (L$_S$) range of 49°-100° of Martian year (MY) 36. Two gaps exist at L$_S$=51°-57° and 93°-97° due to spacecraft safe mode events, so the continuous data at L$_S$=60°-90° are selected to avoid possible influence of these gaps (Figure 1a). This is a dust-clear season in the late northern spring when Mars is near aphelion, and there are no significant seasonal variations of daily temperature anomalies (Text S1, Figures S1-S3). Observations within this 30° of L$_S$ are considered together to improve statistics. The selected temperature profiles total ~7.0×10$^4$ in number, and have a full coverage in geography and local time, despite a slight asymmetry in latitude with more day time sampling in the north and night time in the south (Figure 1b and 1c). Individual profiles are firstly vertically interpolated to the same pressure grid, which is finer than that in the retrieval, and then binned in longitude, latitude, and local time with grid sizes of 5°, 10°, and 1h, respectively (Figure 1c), in the investigations of zonal and diurnal mean temperature and corresponding daily anomalies (Section 3.1). Each bin is assigned the same weight to reduce the biased local time sampling. Uncertainties of the binning include retrieval uncertainties and the variance of retrieved temperatures, and those of zonal and diurnal averaging are computed through error propagation. They are usually small and negligible after binning and averaging (therefore not shown), except for the case of detecting the ter-diurnal tide (Section 3.2).

## 2.3  Wave Mode Decomposition

Contributions of atmospheric waves, including amplitudes (A) and phases (θ), on the diurnal temperature variations are derived using least-square fit with a linear assumption (Gierasch & Goody 1968, Zurek 1976).

$$T(\lambda, \varphi, p, t) = \sum_{\sigma,s} A_{\sigma,s}(\varphi, p) \sin\left(s\lambda + \sigma t + \theta_{\sigma,s}(\varphi, p)\right) \qquad (1)$$

where λ, φ, and p are longitude, latitude, and pressure level, respectively; t is the universal time; s and σ are the wave frequencies in longitude and time. Data are binned only in latitude and interpolated in pressure in this decomposition analysis, as the longitude and time are considered directly. Pairs of the frequencies, (s, σ), denote the wave modes; e.g., (s, σ)=(1, 1) represents the mode with wavenumber one in longitude and a period of one Martian day in time, which is the diurnal tide. Among them, migrating thermal tides propagate westward Sun-synchronously with s/σ=1. Details of the linear regression and the derivation of uncertainties are given in the Supporting Information (Text S2). All wave modes with s={0, 1, 2, 3} and σ={-2, -1, 0, 1, 2} are considered, while σ=±3 is later included due to a visible ter-diurnal tide structure in the residual (Section 3.2).

## 3 Results

### 3.1  Diurnal Temperature Variation

Temperature profiles and estimated retrieval uncertainties obtained near local times of 9h and 21h in the equatorial region are shown in Figure 1d as examples. Consistent differences exist between these two local times. The atmosphere at 21h is colder at ~100Pa, but warmer near the surface and at <10Pa, which is an indication of vertically propagating thermal tides. These profiles are smoother than those obtained from limb sounding, e.g., ~5km of MCS observations, half that of EMIRS (Lee et al. 2009), due to the information content of the near-nadir observations (Figure 1e, Smith et al. 2022). The derived temperature at a certain pressure level is a weighted average of its neighboring pressure levels, so the oscillations in the profiles and therefore the inferred tide amplitudes are smaller.

Zonal and diurnal mean temperature (Figure 1f) is obtained by averaging the binned temperature profiles along the axes of longitude and local time. The temperature structure shows typical solstice features with a warm summer pole and a warming structure at a few to tens of Pa towards the winter pole, which is a result of the downwelling branch of the Hadley circulation. Comparison with the predictions of the Mars Planetary Climate Model (PCM, Forget et al. 1999, Madeleine et al. 2011, Navarro et al. 2014, Forget et al. 2022), where microphysics of radiatively active water clouds are included, is given in the Supporting Information (Text S3, Figure S4). The model generally agrees well with the observation except for some temperature overestimates at low latitudes by a few K, and underestimates near the poles (Figure S4f).

Daily temperature anomalies (Figure 2) are derived by subtracting the zonal and diurnal mean from the zonally averaged binned profiles. This is the first time that such variations are observed on a global scale without any significant gaps in local time or sampling bias in season. The daily anomalies at low latitudes between ±20° (Figure 2f-2j) show signatures of dominant downward phase propagation of diurnal tide with an amplitude of ~6K at <10Pa to ~2K at >100Pa. The temperature maximum propagates approximately from 23h at 5Pa to 19h at 500Pa. At mid-latitudes, however, a large day-night contrast of ~4K extends from surface to ~10Pa at fixed local times (Figure 2c-2e and 2k-2m). A tide-like structure exists at small pressure levels with a temperature anomaly propagating from approximately 8h at 5Pa to 18h at 20Pa in the north (Figure 2k-2m), while it is not clear in the south (Figure 2c-2e). Such a phase transition likely results from a rapid decrease of dust loading near the dust top (Wu et al. 2021), which is also north-south asymmetric due to the topography and its induced meridional circulation (Richardson & Wilson 2002). The derived temperature anomalies at high-latitudes have gaps in local time (Figure 2a-2b and 2n-2o) due to under sampling (Section 2.2, Figure 1c).

### 3.2 Migrating Thermal Tides

Analysis in this work mainly focuses on migrating thermal tides, as they constitute the main diurnal temperature variation in the Martian atmosphere below 60 km (Banfield et al. 2003, Lee et al. 2009., Fan et al. 2022). By applying the least-square fit of Equation (1) to the observed temperatures (Section 2.3), amplitudes and phases of the tides are derived. The result in the equatorial bin between ±5° is shown in Figure 3. Combination of the modes with the time frequency, $\sigma$, truncated at ±2 (Figure 3a) reproduces most of the diurnal temperature variation (Figure 2h), with residual less than 0.8K at most pressure levels (Figure 3b). However, the residual is consistently larger than the uncertainty, and with patterns of downward phase progression with a period of one third of a Martian day appear (Figure 3b), which suggests the existence of ter-diurnal tide. Therefore, the wave mode decomposition is then reapplied with $\sigma$ expanded to ±3. Although the resulting diurnal temperature variation does not change much (Figure 3c), the residual is mostly below the uncertainty level and becomes random (Figure 3d). The inclusion of ter-diurnal tide greatly improves the decomposition, which indicates its existence.

Contributions of the first three migrating thermal tides and their phases are shown in Figure 4 for the same equatorial bin. The diurnal tide, the (1, 1) mode, dominates the diurnal temperature variation with an amplitude of ~2-6K (Figure 4a). Its phase progression is linear with the logarithm of pressure (Figure 4b), which suggests a constant wavelength of ~40km if assuming a scale height of 10km. Compared to the model predictions, the observed diurnal tide has a similar vertical wavelength, but an earlier phase of ~3h. Both of the observed and modeled wavelengths become larger at pressure levels <20Pa, which is likely due to the difference in zonal wind and/or excitation sources of dust/clouds (Wu et al. 2017). The semi-diurnal tide, the (2, 2) mode, shows an amplitude of ~1.5-2K across all pressure levels (Figure 4c). Its phase progression is also linear, and indicates a wavelength of ~200km (Figure 4d), which is far larger than that in the model prediction (~60km for the original output, or ~80km if sampling and vertical convolution are considered). Such a large wavelength indicates a possible dominant trapped Hough mode, which does not vertically propagate outside the region of excitation sources (likely water ice clouds in the aphelion tropics; Kleinböhl et al. 2013, Wilson et al. 2014, Haberle et al. 2020). As a new finding, the ter-diurnal tide, the (3, 3) mode, has an amplitude of ~0.3-0.5K (Figure 4e), which is well above the uncertainty level at ~30-200Pa where the retrieved temperatures are best constrained (Figure 3d). The inferred phase agrees with the model (different by <1h) at low pressure levels (~5-50Pa), but it is completely different at hundreds of Pa (Figure 4f), which corresponds to the three temperature maxima at ~4, ~12, and ~20h shown in the residual (Figure 3b). This wave mode may result from the wavenumber 3 of subtropic topography, but approval or negation requires future numerical simulations.

Latitudinal and vertical distributions of amplitudes and phases of the migrating tides are derived by repeating the wave mode decomposition in each latitude bin and at pressure level (Figure 5). The diurnal tide has a maximal amplitude of ~6K near the equator at ~5Pa, and also large values north of 30°N (Figure 5a). Such a latitudinal distribution agrees with the dominant propagating (1, 1) Hough mode. The phase progression of the diurnal tide is well constrained in the equatorial region between ±20°, while it has a constant value across a range of pressure levels at mid-latitudes (Figure 5b). This constant phase corresponds to the vertically extended day-night temperature contrast (Figure 2k-2m), and indicates trapped Hough modes in subtropics. Similar to that in the equatorial bin (Figure 4c-4d), the semi-diurnal tide between ±20° has an amplitude of ~2K (Figure 5c), but with slightly different downward propagating phases (Figure 5d). The phase in the northern hemisphere is earlier than that in the south, which is likely due to asymmetric dust loading or cloud extension caused by topography-induced meridional circulation (Richardson & Wilson 2002). The ter-diurnal tide has a maximal amplitude of ~0.5K at ~20Pa (Figure 5e), and also a downward phase progression at most latitudes (Figure 5f). Its phase distribution seems to have a symmetrical pattern about 20°N, which serves as a reference for further investigations.

## 4 Discussion and Conclusion

Diurnal temperature variations in the Martian atmosphere are investigated at $L_S$=60°-90° of MY 36, using temperature profiles retrieved from the first set of EMM/EMIRS observations. The data show a dominant diurnal tide and an important semi-diurnal tide, as well as the existence of ter-diurnal tide. Compared to the Mars PCM, all migrating tides show similar amplitudes providing that the coarse resolution of EMIRS is taken into account, but the observed diurnal tide has an earlier phase, and the wavelength of the semi-diurnal tide is unexpectedly large.

Due to the novel and high-altitude design of the spacecraft orbit, EMM/EMIRS observes diurnal temperature variations in the Martian atmosphere on a global scale, with all location and local time covered within a short range of season. This coverage is essential for detailed analysis of thermal tides, and was one of the major issues in previous works. Observations obtained by TES and MCS on Sun-synchronous orbits are usually around two local times separated by half of a Martian day (Banfield et al. 2003, Lee et al. 2009), which contain strong aliasing of wave modes, even with unequally-spaced cross-track observations (±1.5-3.0h) included (Kleinböhl et al. 2013, Wu et al. 2015, 2017). Dealiasing of these wave modes requires significant assumptions of the Martian atmospheric physical properties and knowledge from tidal theory (Lindzen & Chapman 1969). Observations obtained by PFS and TIRVIM on slowly drifting orbits (Giuranna et al. 2021, Fan et al. 2022) have strong seasonal change influence in the interpretation of diurnal variations. The advantage of EMM/EMIRS observations that cover all geographic locations and local times within 10 days (~5° in $L_S$) largely addresses this issue, which enables detailed tide investigations with good constraints on their amplitudes and phases, as well as the detection of the ter-diurnal tide. Effects of the observational scheme including sampling and vertical convolution on the tide interpretations are shown in the Supporting Information (Text S3). Sampling does not make noticeable influence and is no longer an issue; vertical convolution decreases the interpreted amplitude by a factor of ~2 and results in smaller phase changes. Amplitudes of diurnal and semi-diurnal tides agree well with results in previous works during this aphelion season (Banfield et al. 2003, Kleinböhl et al. 2013, Fan et al. 2022), but the phases show significant differences. The inferred amplitude of the ter-diurnal (~0.3-0.5K) is also consistent with the TIRVIM results (<0.5K, Fan et al. 2022), where the data sampling scheme was not sufficiently good for detecting this mode.

Amplitudes and phases of thermal tides are usually indicators of their excitation sources, among which airborne dust and water ice clouds are two key factors (Hinson & Wilson 2004, Guzewich et al., 2013, Kleinböhl et al. 2013, Wilson & Guzewich, 2014, Wu et al., 2017, 2021). During the dust-clear aphelion season, water ice clouds play an important role in shaping the temperature structure of the Martian atmosphere (Wilson et al. 2008, Wilson et al. 2014), and are among major sources exciting diurnal and semi-diurnal tides (Kleinböhl et al. 2013, Wilson et al. 2014, Haberle et al. 2020). Disagreements between observations and model predictions shown in this analysis suggest improvements in numerical simulations. The earlier phase of the diurnal tide and the likely trapped modes of both diurnal and semi-diurnal tides provide constraints on the vertical distribution of dust and/or clouds as well as their particle sizes and radiative processes. New mechanisms are needed to explain the excitation and distribution of the ter-diurnal tide. These are important in enriching our understanding of the Martian atmosphere on a diurnal basis.

### Acknowledgments

Funding for development of the EMM mission was provided by the United Arab Emirates (UAE) government, and to co-authors outside of the UAE by the Mohammed bin Rashid Space Centre (MBRSC). RMBY acknowledges funding from UAE University grants G00003322 and G00003407.

### Open Research

Data from the Emirates Mars Mission (EMM) are freely and publicly available on the EMM Science Data Center (SDC, http://sdc.emiratesmarsmission.ae). This location is designated as the primary repository for all data products produced by the EMM team and is designated as long-term repository as required by the UAE Space Agency. The data available (http://sdc.emiratesmarsmission.ae/data) include ancillary spacecraft data, instrument telemetry, Level 1 (raw instrument data) to Level 3 (derived science products), quicklook products, and data users guides (https://sdc.emiratesmarsmission.ae/documentation) to assist in the analysis of the data. Following the creation of a free login, all EMM data are searchable via parameters such as product file name, solar longitude, acquisition time, sub-spacecraft latitude & longitude, instrument, data product level, and etc. Emirates Mars Infrared Spectrometer (EMIRS) data and users guides are available at: https://sdc.emiratesmarsmission.ae/data/emirs.

Data products can be browsed within the SDC via a standardized file system structure that follows the convention:

/emm/data/<Instrument>/<DataLevel>/<Mode>/<Year>/<Month>

Data product filenames follow a standard convention: emm_<Instrument>_<DataLevel><StartTimeUTC>_<OrbitNumber>_<Mode>_<Description>_<KernelLevel>_<Version>.<FileType>

The Mars PCM output during Ls=0°-90° of MY 36 is available on the IPSL data server with doi: 10.14768/d49ef040-476c-4264-bf67-6b4b018b8620. Permission is granted to use these datasets in research and publications with appropriate acknowledgements that are presented on the dataset websites.

**References from the Supporting Information**

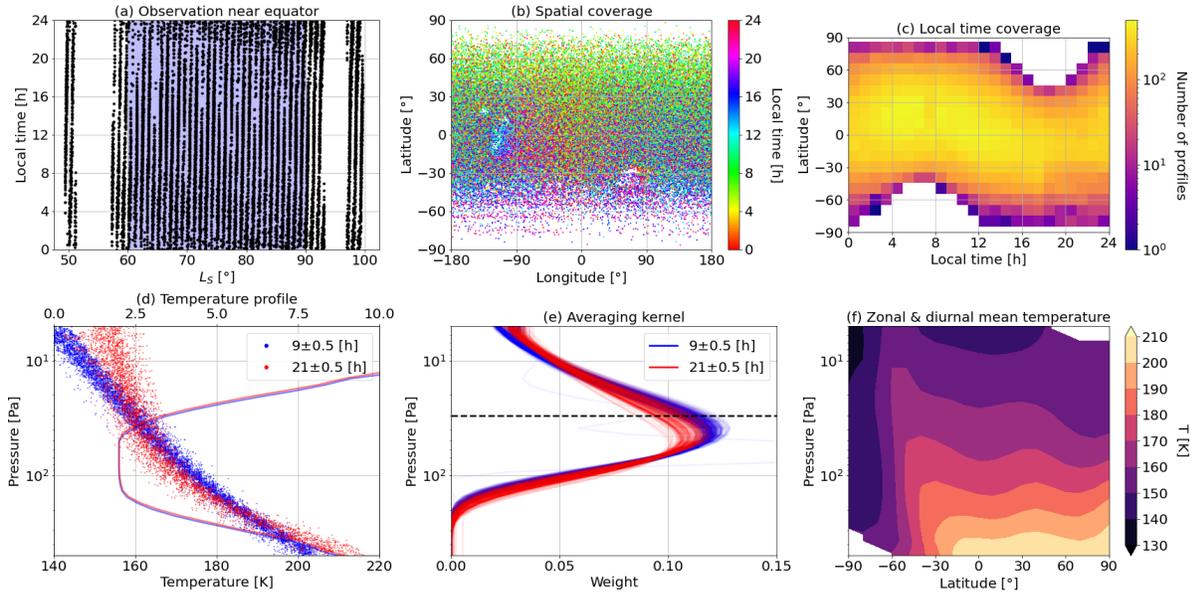

**Figure 1.** (a) Season and local time of the first set of EMIRS temperature profile observations between ±5° in latitude (back dots). The blue shaded area denotes the season ($L_S$=60°-90°) selected in this work. (b) Location of the selected EMIRS observations. The color denotes the observation local time. (c) Number of EMIRS observations in the (latitude, local time) bins. (d) Temperatures (color dots, lower axis) and estimated uncertainties (color lines, upper axis) retrieved using EMIRS observations between ±5° in latitude and within 0.5h of 9h (blue) and 21h (red). (e) Averaging kernels of retrieving the temperatures at 30Pa (black dashed line) using observations between ±5° in latitude and within 0.5h of 9h (blue lines) and 21h (red lines). The retrieved pressure levels are averages weighted by corresponding asymmetric kernels, so they do not coincide with maximal values of the kernels. (f) Zonal and diurnal mean temperature derived using EMIRS observations during MY 36 $L_S$=60°-90°.

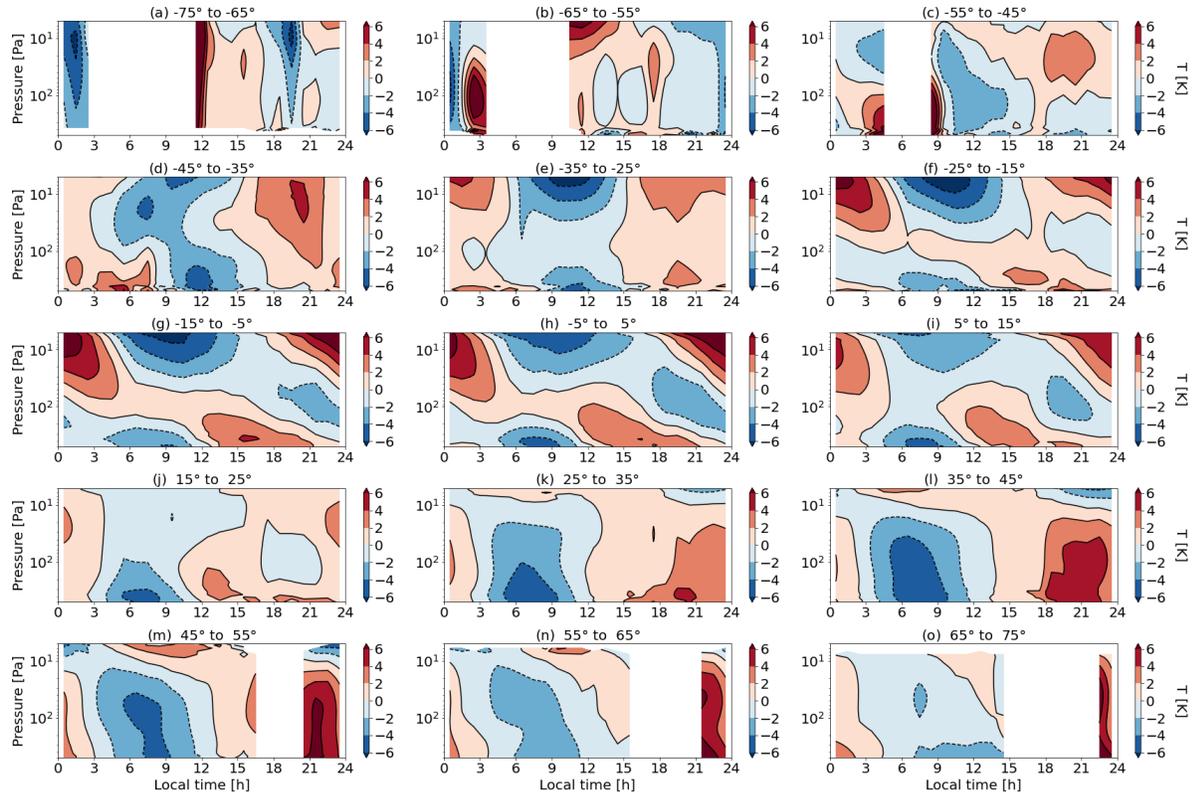

**Figure 2.** Zonal mean daily temperature anomalies derived using EMIRS observations for latitude bins centered at 70°S to 70°N with an interval of 10°.

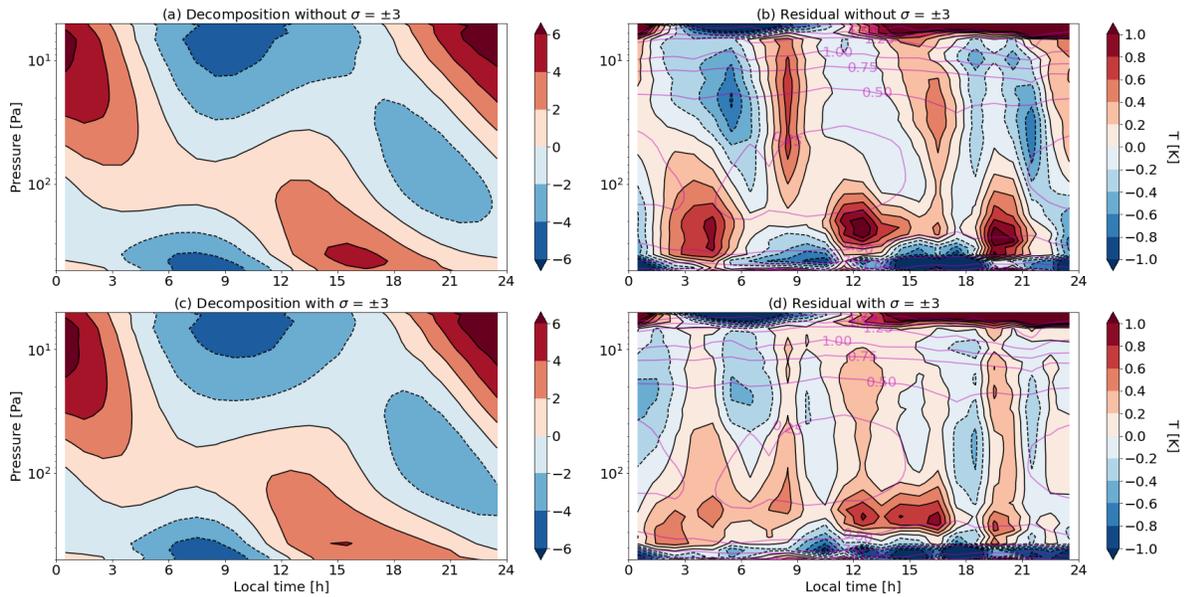

**Figure 3.** (a) Wave mode decomposition result of the daily temperature anomaly in the equatorial bin between ±5° with the time frequency, σ, truncated at ±2. (b) Residual (filled contours) of the wave mode decomposition shown in (a), which is its difference from Figure 2h, and the combined uncertainty (magenta contour lines), which includes that from both observation and wave mode decomposition. The interval of uncertainty levels is 0.25K. (c) and (d) Same as (a) and (b), respectively, but for decomposition with σ=±3 included.

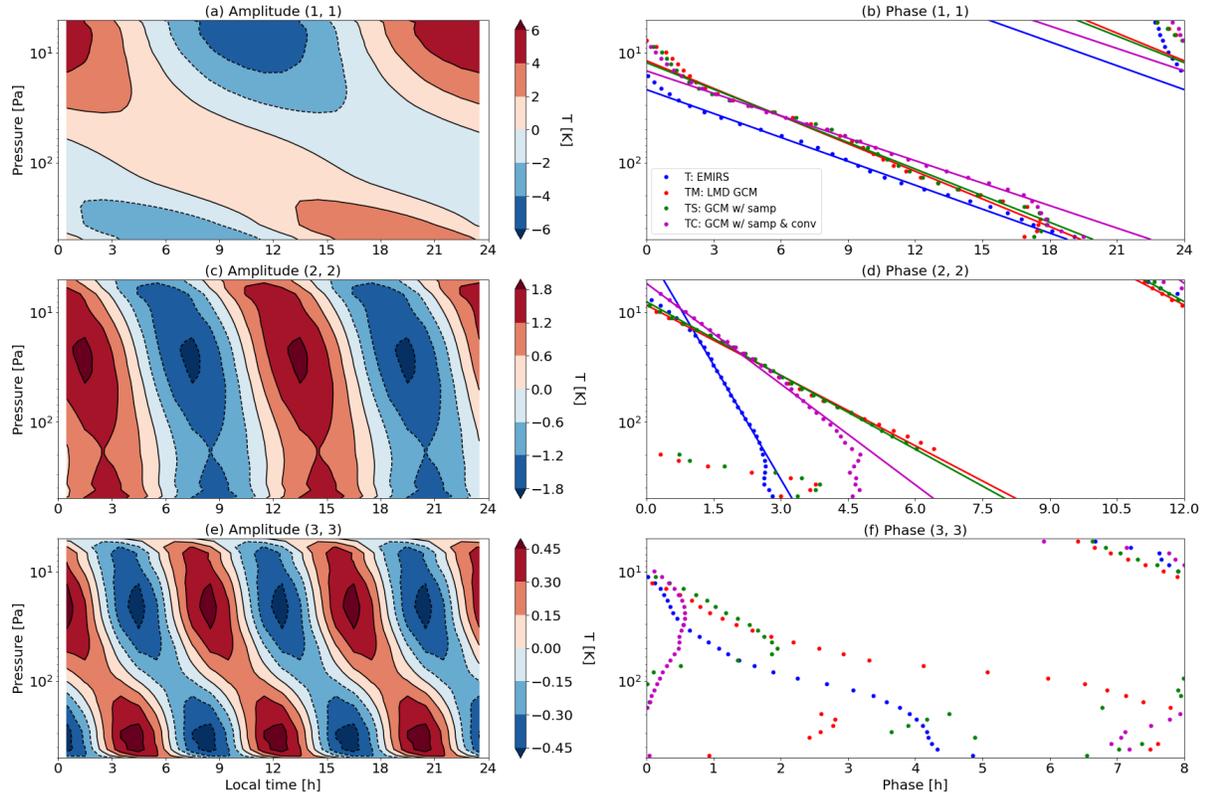

**Figure 4.** (a) Diurnal thermal tide derived from the wave mode decomposition using EMIRS observations in the equatorial bin between ±5°. (b) Phase of the diurnal tide, represented by the local time of the temperature maximum, in the equatorial bin derived using EMIRS observations (blue dots), the Mars PCM outputs (red dots), the model outputs sampled at the same locations and times with observations (green dots), and the sampled model outputs with vertical convolution included (magenta dots). The color lines denote the approximate linear downward phase progressions. (c) and (d) Same as (a) and (b), respectively, but for the semi-diurnal tide. (e) and (f) Same as (a) and (b), respectively, but for the ter-diurnal tide.

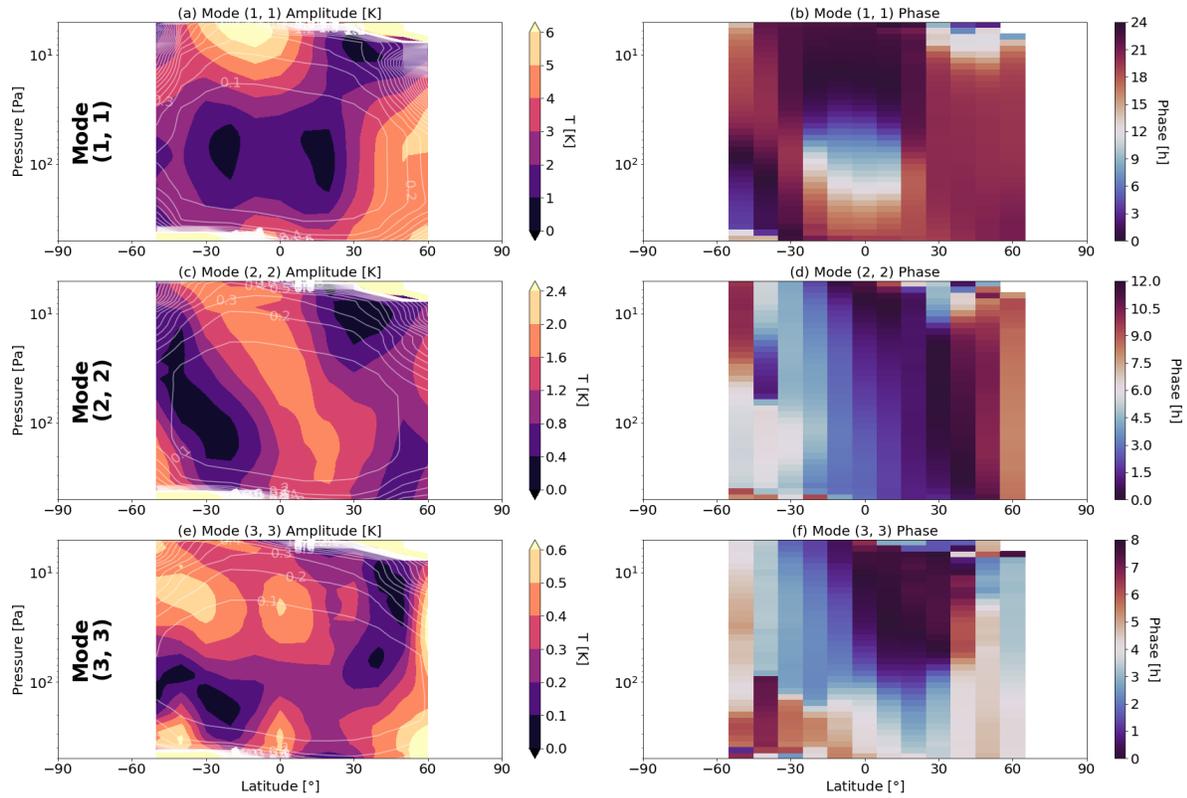

**Figure 5.** (a) Amplitude (filled contours) and uncertainty (while contour lines) of the diurnal tide component derived using EMIRS observations. The interval of the uncertainty level is 0.1K. (b) Same as (a), but for the phase of the diurnal tide, denoted by the local time of the temperature maximum. (c) and (d) Same as (a) and (b), respectively, but for the semi-diurnal tide. (e) and (f) Same as (a) and (b), respectively, but for the ter-diurnal tide.

*Geophysical Research Letters*

Supporting Information for

**Migrating Thermal Tides in the Martian Atmosphere**

**during Aphelion Season Observed by EMM/EMIRS**

Siteng Fan[1], François Forget[1], Michael D. Smith[2], Sandrine Guerlet[1], Khalid M. Badri[3], Samuel A. Atwood[4,5], Roland M. B. Young[6], Christopher S. Edwards[7], Philip R. Christensen[8], Justin Deighan[5], Hessa R. Al Matroushi[3], Antoine Bierjon[1], Jiandong Liu[1], Ehouarn Millour[1]

[1]LMD/IPSL, Sorbonne Université, PSL Research Université, École Normale Supérieure, École Polytechnique, CNRS, Paris, France
[2]NASA Goddard Space Flight Center, Greenbelt, MD, USA
[3]Mohammed Bin Rashid Space Centre, Dubai, UAE
[4]Space and Planetary Science Center, and Department of Earth Sciences, Khalifa University, Abu Dhabi, UAE
[5]Laboratory for Atmospheric and Space Physics, Boulder, CO, USA
[6]Department of Physics & National Space Science and Technology Center, United Arab Emirates University, Al Ain, UAE
[6]Department of Astronomy and Planetary Science, Northern Arizona University, Flagstaff, AZ, USA7
[8]School of Earth and Space Exploration, Arizona State University, Tempe, AZ, USA

**Contents of this file**

> Texts S1 to S3
> Figures S1 to S9

## Introduction

This supporting information includes the following content.

Text S1: Diurnal temperature anomalies derived using EMIRS observations binned every 5° in $L_S$, which shows negligible seasonal changes in the range of $L_S$=60-90°.

Text S2: Details of the wave mode decomposition and the derivation of corresponding uncertainty.

Text S2: Description of the Mars PCM, which is used for simulating the Martian atmosphere in this work and serves as a baseline of our current understanding.

Figures S1-S3: Zonal mean daily temperature anomalies derived using EMIRS observations binned every 5° in $L_S$

Figure S4: Zonal and diurnal mean temperature in the Martian atmosphere during MY 36 $L_S$=60°-90° computed using EMIRS observations and the Mars PCM outputs, also shown with their comparisons.

Figure S5: Zonal mean daily temperature anomalies computed using the sampled and convolved model outputs.

Figure S6: Zonal mean daily temperature anomaly at latitude between ±5°, computed using EMIRS observations and the Mars PCM outputs, also shown with their comparisons.

Figures S7-S9: Amplitudes and phases of the first three migrating tides, computed using the model outputs.

**Text S1.** Seasonal variation

     Diurnal temperature anomalies binned within different $L_S$ ranges between 60 and 90° are shown in Figure S1-S3 for the equatorial region and mid-latitudes. The data are interpolated onto the same pressure levels and binned using the same grid in longitude, latitude, and local time (Section 2.2), except that the selected $L_S$ range is every 5°, which is the minimal required time for EMIRS observations to have a full coverage in geography and local time. All of the diurnal temperature anomalies show similar features within different $L_S$ ranges, and the differences among them are mostly random and below the uncertainty level. There is no clear trend or consistent seasonal changes identified in the diurnal temperature anomalies. Therefore, for the purpose of improving the statistics and reducing the random noise, all data within $L_S$=60-90° are binned together in this work for the analyses of diurnal temperature variations and thermal tides in the Martian atmosphere.

**Text S2.** Wave mode decomposition

The least-square fit with a linear assumption has been shown successful in analyzing observations of many previous missions, e.g., MCS (Wu et al. 2015, 2017) and TIRVIM (Fan et al. 2022), despite that the unequally-spacing observations centered around two local times of MCS result in large uncertainty of the interpreted semi-diurnal tide, and that the slow local time drifting of TIRVIM data introduce seasonal variations into the diurnal analysis. In a given latitude bin (φ) and at a certain pressure level (p), the observed temperature (T) is assumed to be a function of longitude (λ) and universal time (t), so Equation (1) can be rewritten to achieve the linearity of unknowns.

$$T(\lambda,\ t) = \sum_{\sigma,s}\big(C_{\sigma,s}\cos(s\lambda + \sigma t) + S_{\sigma,s}\sin(s\lambda + \sigma t)\big) \qquad (2)$$

where C and S are the coefficients of the Fourier bases of the wave modes at this latitude and pressure level. Their amplitudes (A) and phases (θ) have the following relationships with the coefficients.

$$A_{\sigma,s} = \sqrt{C_{\sigma,s}{}^2 + S_{\sigma,s}{}^2} \qquad (3)$$

$$\theta_{\sigma,s} = \tan^{-1}\left(\frac{C_{\sigma,s}}{S_{\sigma,s}}\right) \qquad (4)$$

A linear regression problem is reached by rearranging Equation (2) and considering all temperature observations in this latitude bin and at this pressure level.

$$\mathbf{W}_{[N\times D]}\boldsymbol{x}_{[D\times 1]} + \mathbf{e}_{[N\times 1]} = \boldsymbol{y}_{[N\times 1]} \qquad (5)$$

where **x, y, e** are column vectors of the coefficients (C and S), temperature observations, and temperature uncertainties, respectively; **W** is the weight matrix whose (i, j)-th element is the value of the j-th Fourier basis, either a cosine or a sine function, of the i-th observation, which gives λ and t; N is the total number of observations, which is usually on the order of thousands depending on the selected latitude bin; D is the total number of linearly independent coefficients, which is 35 when s={0, 1, 2, 3} and σ={-2, -1, 0, 1, 2}, and 53 when σ is expanded to {-3, -2, -1, 0, 1, 2, 3}. Finally, this linear regression problem can be solved using least-square fit, and the expected value and covariance matrix of **x** can be derived as follows.

$$\mathbf{E}[\boldsymbol{x}] = (\mathbf{W^T C^{-1} W})^{-1}(\mathbf{W^T C^{-1}})\mathbf{y} \qquad (6)$$

$$\mathbf{Cov}[\boldsymbol{x}] = [(\mathbf{W^T C^{-1} W})^{-1}(\mathbf{W^T C^{-1}})\boldsymbol{e}][(\mathbf{W^T C^{-1} W})^{-1}(\mathbf{W^T C^{-1}})\boldsymbol{e}]^{\mathrm{T}}$$
$$= (\mathbf{W^T C^{-1} W})^{-1} \qquad (7)$$

where **C** is the covariance matrix of observations, which is a diagonal matrix with the (i, i)-th element to be the square of the i-th element of **e**, as the observations are independent from each other. Therefore, the uncertainty of the i-th element of **x**, which is one of the coefficients (C or S) by definition, is the square root of the (i, i)-th element of **Cov[x]**, and the uncertainties of all wave mode amplitudes can be derived through error propagation using Equation (3).

**Text S3.** Mars PCM

Serving as a baseline of our current understanding of the Martian atmosphere, predictions from the Mars Planetary Climate Model (PCM), which is previously known as the Laboratoire de Météorologie Dynamique Mars Global Circulation Model (LMD Mars GCM, version 5, Forget et al. 1999), is used for the model-observation comparison. This model is one of the most popular Mars GCMs, which has been updated for decades (e.g., Madeleine et al. 2011, Navarro et al. 2014, Forget et al. 2022). The Mars PCM computes atmospheric dynamics in three dimensions, while the forcings of physical processes are derived in each vertical column. Horizontal resolutions of the model are 5.625° and 3.75° in longitude and latitude, respectively, and the vertical grid contains 32 layers from the Martian surface to ~$2 \times 10^{-3}$Pa, without the inclusion of the Martian thermosphere. The model is set to have 960 steps in each Martian day for computing the dynamics, and the physical forcings are updated every 10 steps. The physical processes include (1) radiative processes of $CO_2$, dust, and clouds, (2) $CO_2$, dust, and water cycles, and (3) transport of dust and gas, etc. The version 5 used in this work includes microphysics of radiatively active water ice clouds, but not rocket dust storms, mountain top flows, or $CO_2$ ice-induced scavenging processes. The column-integrated dust opacity is normalized at each time step to the MY 36 scenario built following Montabone et al. (2020).

Figure S4 shows the model-observation comparison of the zonal and diurnal mean temperature. For the purpose of appropriate comparison with the observed result (T, Figure S4a), the original model results (TM, Figure S4b) are first sampled at the same times and locations with the observations (TS, Figure S4c), and then vertically convolved (TC, Figure S4d) using the averaging kernels derived from the retrieval (Figure 1e, Smith et al. 2022), which is also discussed in Section 3.1. Although the direct comparison shows smaller differences (Figure S4e), it becomes more consistent when the observation scheme is considered (Figure S4f). The model overestimates the zonal and diurnal mean temperature by 2-4K at most pressure levels at low latitudes, but underestimates over poles. This is an indication of improvements in the dust and/or water clouds simulations, which are also suggested by the thermal tide analyses in this work.

The predictions of the daily temperature anomalies are similar to the observations across all latitudes when the observation scheme is considered (Figure 2 and S5). Features of the diurnal temperature variations at different latitudes are also the same, with downward phase-progression structure in the equatorial region between ±20° (Figure S5f-S5j) and dominant day-night temperature contrast at mid to high latitudes (Figure S5c-S5e and S5k-S5m), likely due to trapped Hough modes. However, the phase difference is noticeable in the structure of temperature anomalies (Figure S6). The diurnal thermal tide dominates both the observed and modeled diurnal temperature variations, with a local time difference of ~3h (Figure S6a and S6d), which is also shown in Section 3.2 (Figure 4).

Results of the wave mode decomposition of the first three migrating tides are shown in Figure S7-S9 for the original Mars PCM outputs (TM), the sampled model outputs (TS), and the sampled outputs with vertical convolution (TC). They show that sampling of EMIRS observations is not an issue due to its good data coverage, which was one of the most important factors that prevent the thermal tide analysis using observations from many previous missions (Banfield et al. 2003, Lee et al. 2009, Kleinböhl et al. 2013, Wu et al. 2015, 2017, Fan et al. 2022). Decomposition results of the amplitudes and phases of all three tides do not change before and after the model outputs are sampled (Figure S7a-S7d, S8a-S8d, and S9a-S9d), except for the resolution decrease. Vertical convolution smooths temperature oscillations, so it results in smaller amplitudes (by a factor of ~2) and smoother interpreted phase changes (Figure S7c-S7f, S8c-S8f, and S9c-S9f). The model predictions agree well with the observed diurnal and semi-diurnal tides (Figure 5a-5d, S7, and S8), except for the phase and wavelength differences discussed in Section 3.2. On the contrary, the agreement of ter-diurnal tide is relatively poorer (Figure 5e-5f and S9), which requires further attention.

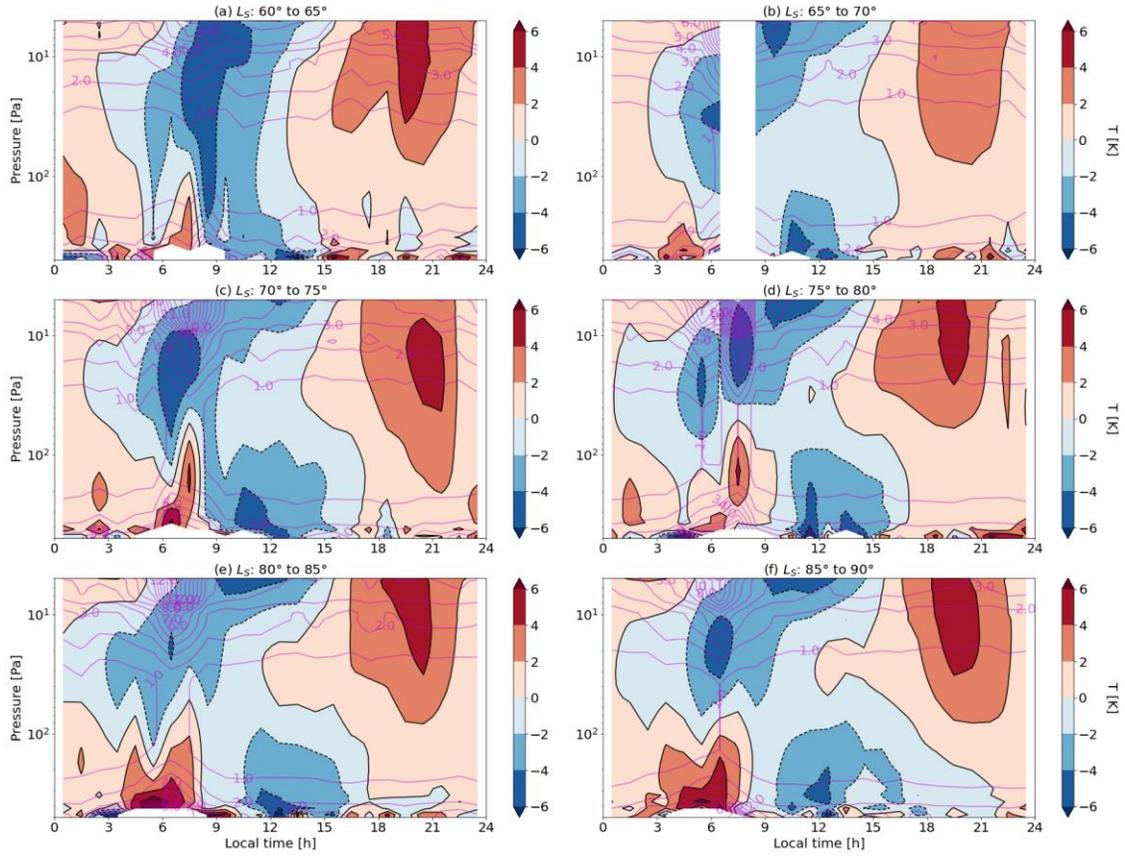

**Figure S1.** Zonal mean daily temperature anomalies (filled contours) and uncertainties (magenta contour lines) derived using EMIRS observations for the latitude bin of 45°S to 35°S, using data binned every 5° in L$_S$ from 60° to 90°. The interval of uncertainty levels is 1K.

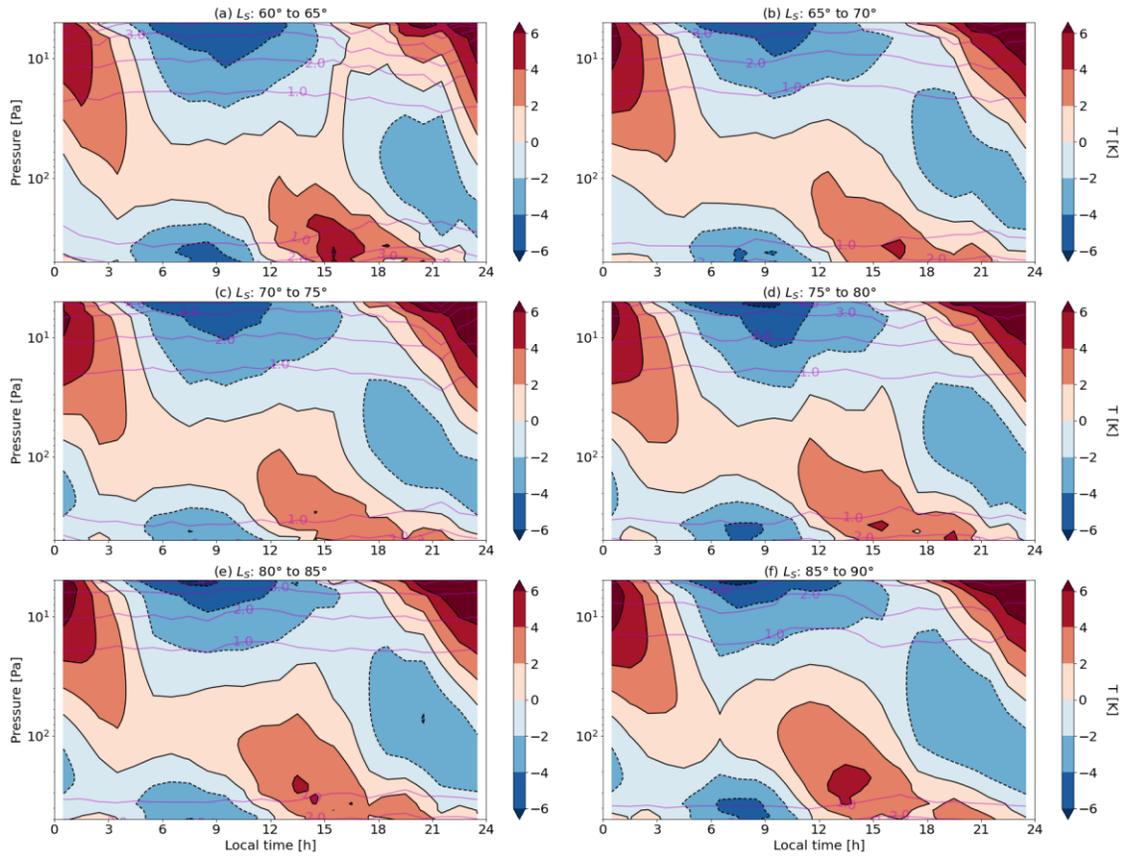

**Figure S2.** Same as Figure S1, but for the latitude bin of 5°S to 5°N.

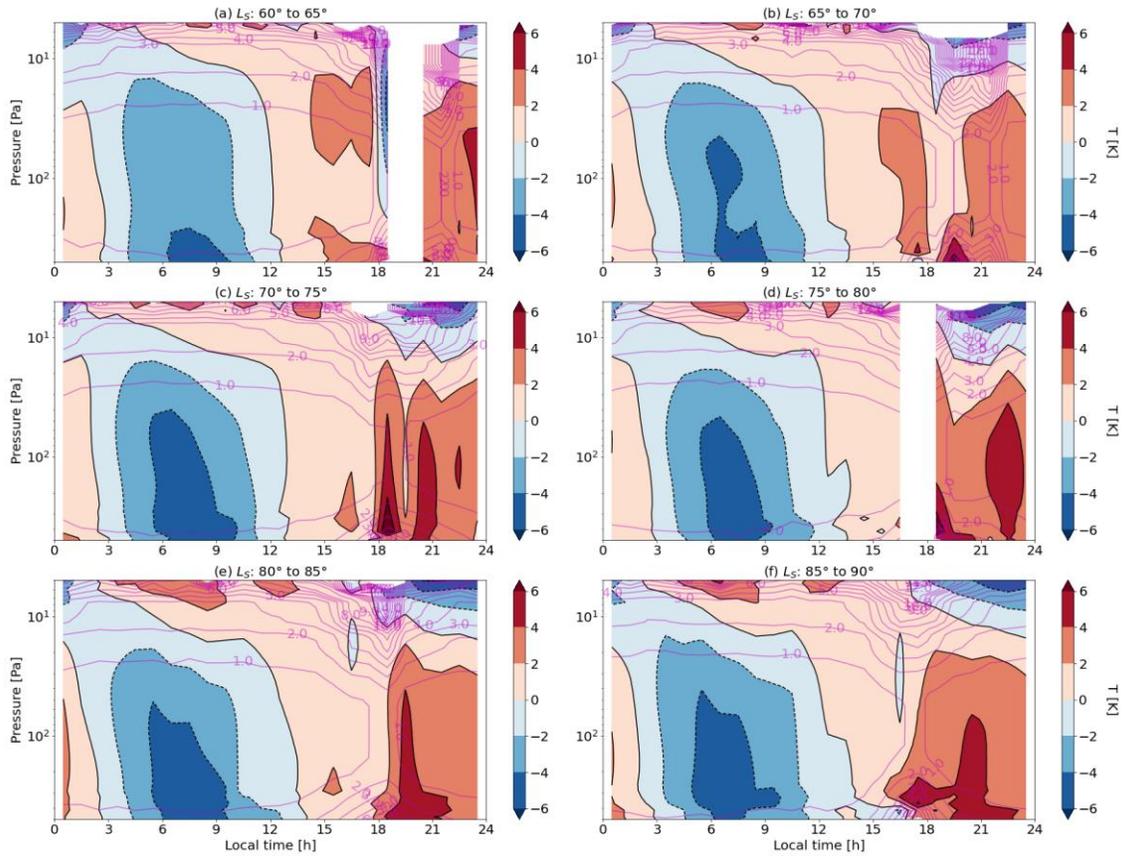

**Figure S3.** Same as Figure S1, but for the latitude bin of 35°N to 45°N.

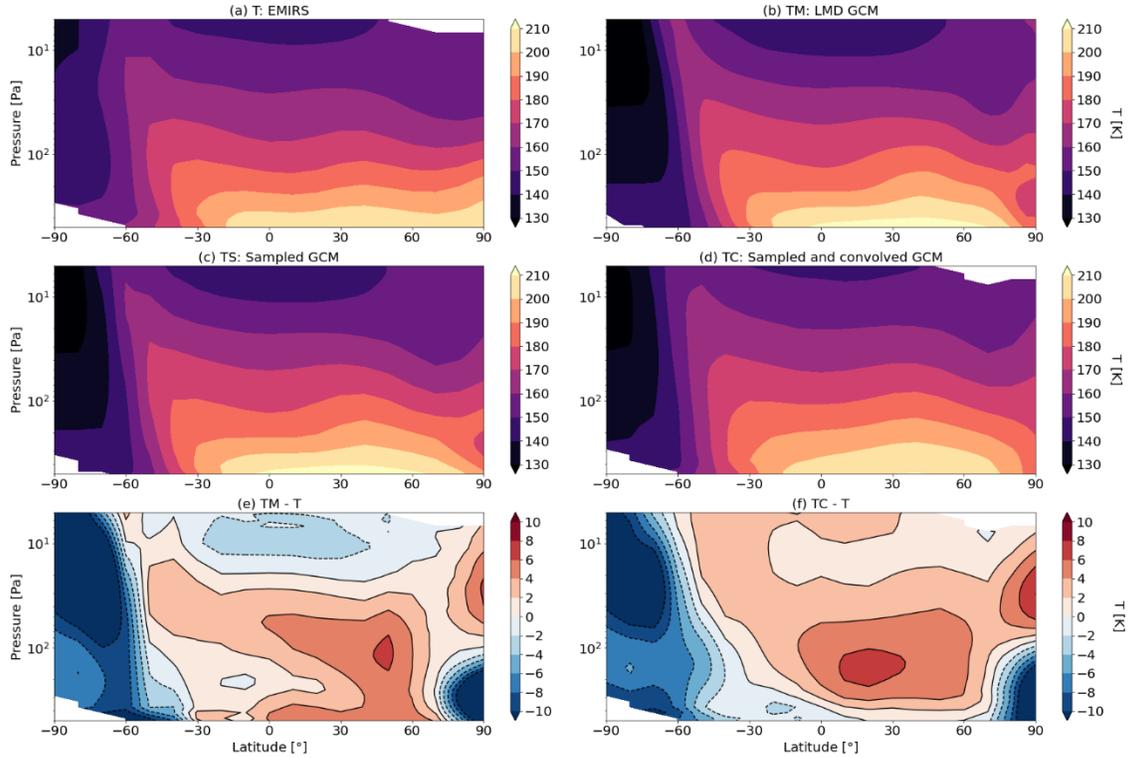

**Figure S4.** (a) Zonal and diurnal mean temperature computed using EMIRS observations during MY 36 $L_S$=60°-90°, same as Figure 1f. (b) Same as (a), but for the Mars PCM outputs. (c) Same as (b), but the outputs are sampled at the same locations and times as EMIRS observations. (d) Same as (c), but it includes the vertical convolution. (e) Difference of the zonal and diurnal mean temperature between observations and the original model outputs, which is the difference between (a) and (b). (f) Same as (e), but for the difference between (a) and (d), where the model outputs are sampled and convolved the same way as observations, which is for appropriate model-observation comparison.

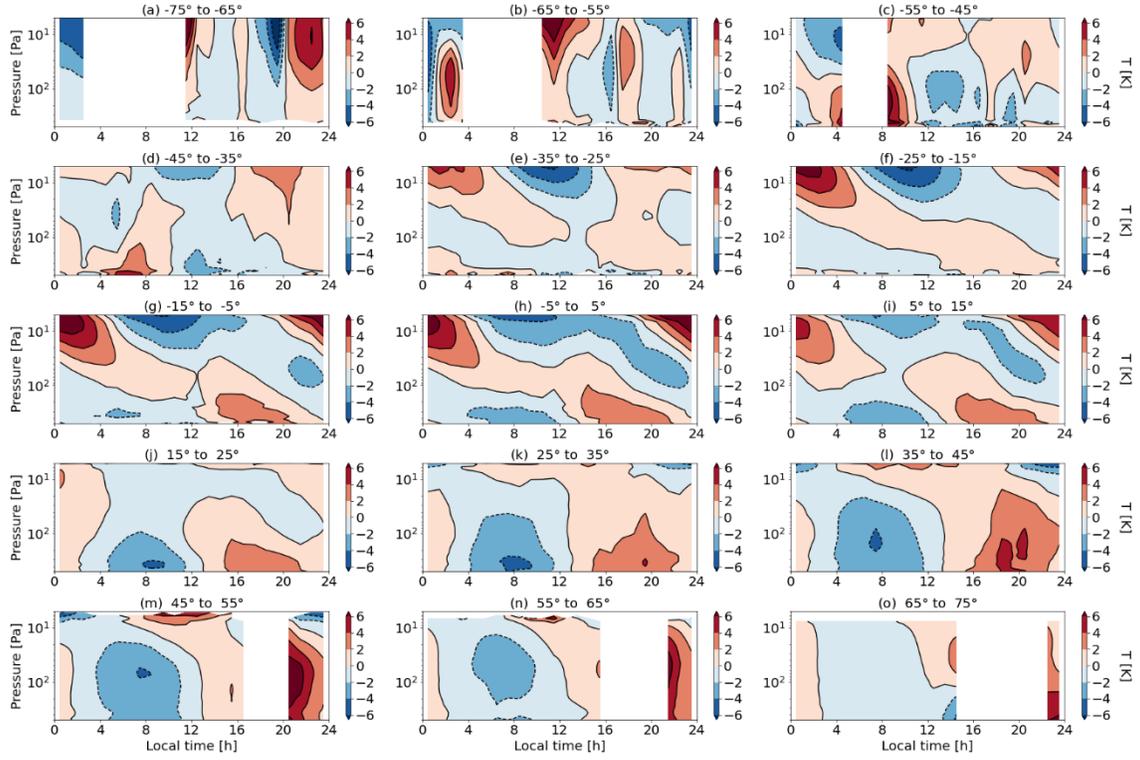

**Figure S5.** Same as Figure 2, but computed using model outputs sampled at the same locations and times as EMIRS observations, and vertically convolved using averaging kernels.

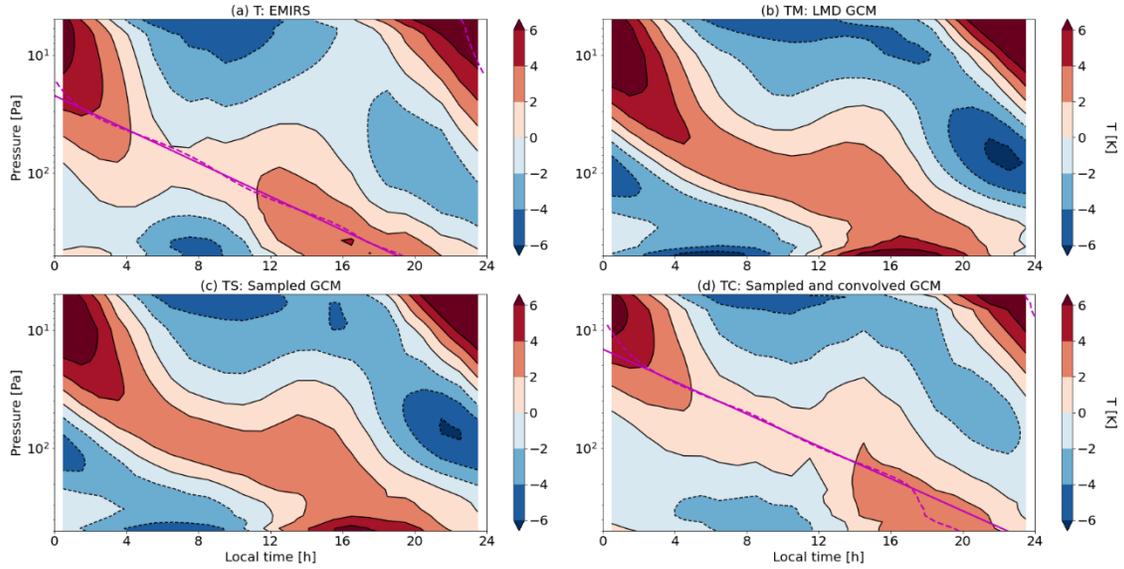

**Figure S6.** Same as Figure S4a-S4d, but for the zonal mean diurnal temperature anomaly at latitude between ±5°. (a) and (d) are the same as Figure 2h and S5h, respectively. The dashed magenta lines denote the phases of the diurnal tide components derived in the wave mode decomposition, and the solid magenta lines are linear fit results representing the approximate tide propagation. The model-observation comparison should be done by comparing (a) and (d).

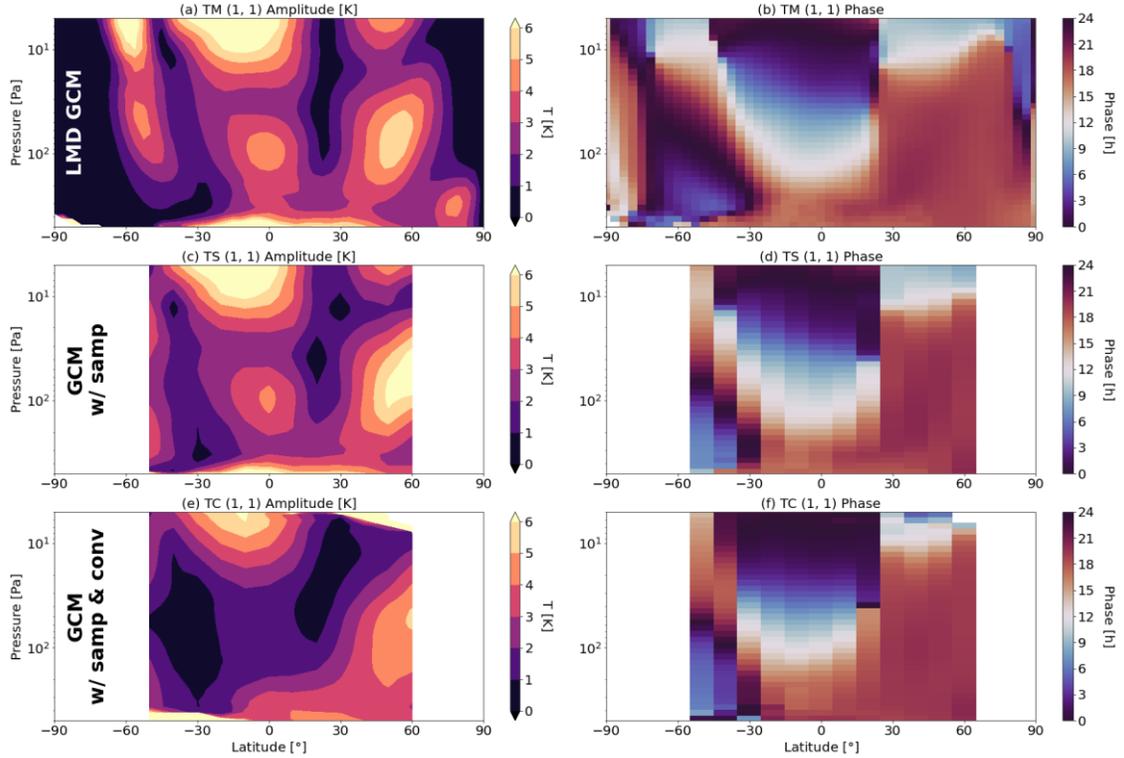

**Figure S7.** (a) and (b) Same as Figure 5a and 5b, respectively, but for the results computed using the Mars PCM outputs. (c) and (d) Same as (a) and (b), respectively, but for the model outputs sampled at the same locations and times as EMIRS observations. (e) and (f) Same as (a) and (b), respectively, but for the sampled model outputs including vertical convolution.

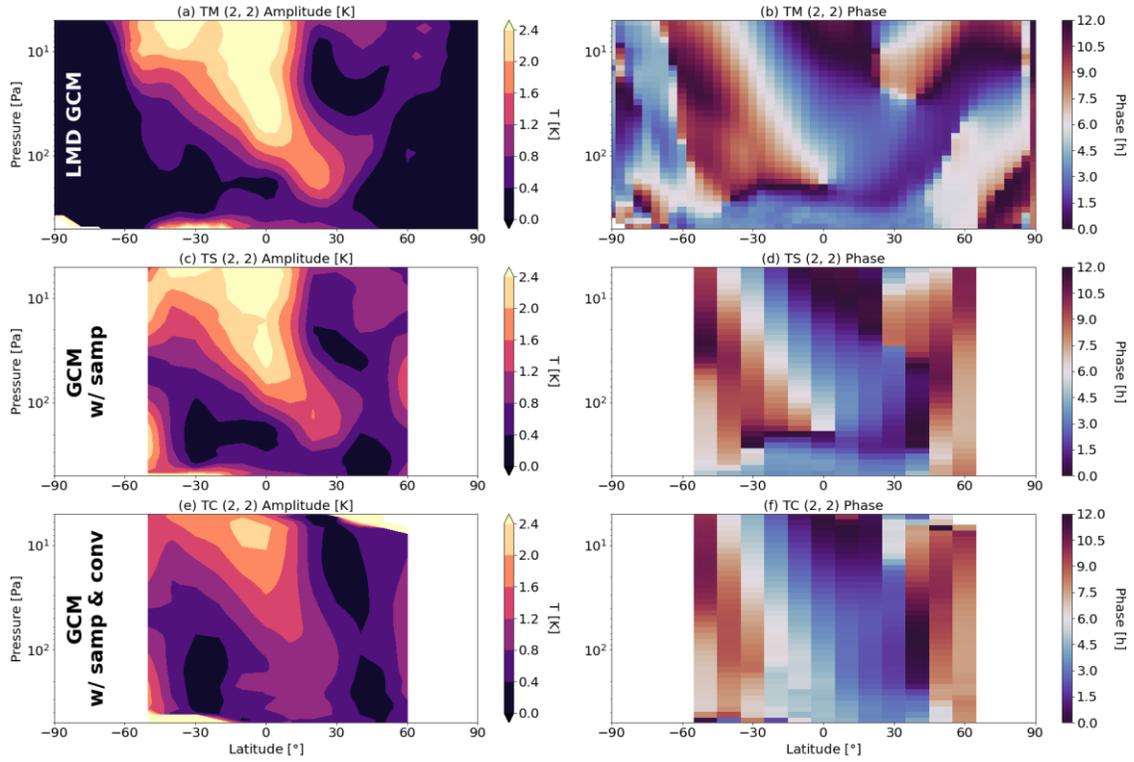

**Figure S8.** Same as Figure S7, but for the semi-diurnal tide.

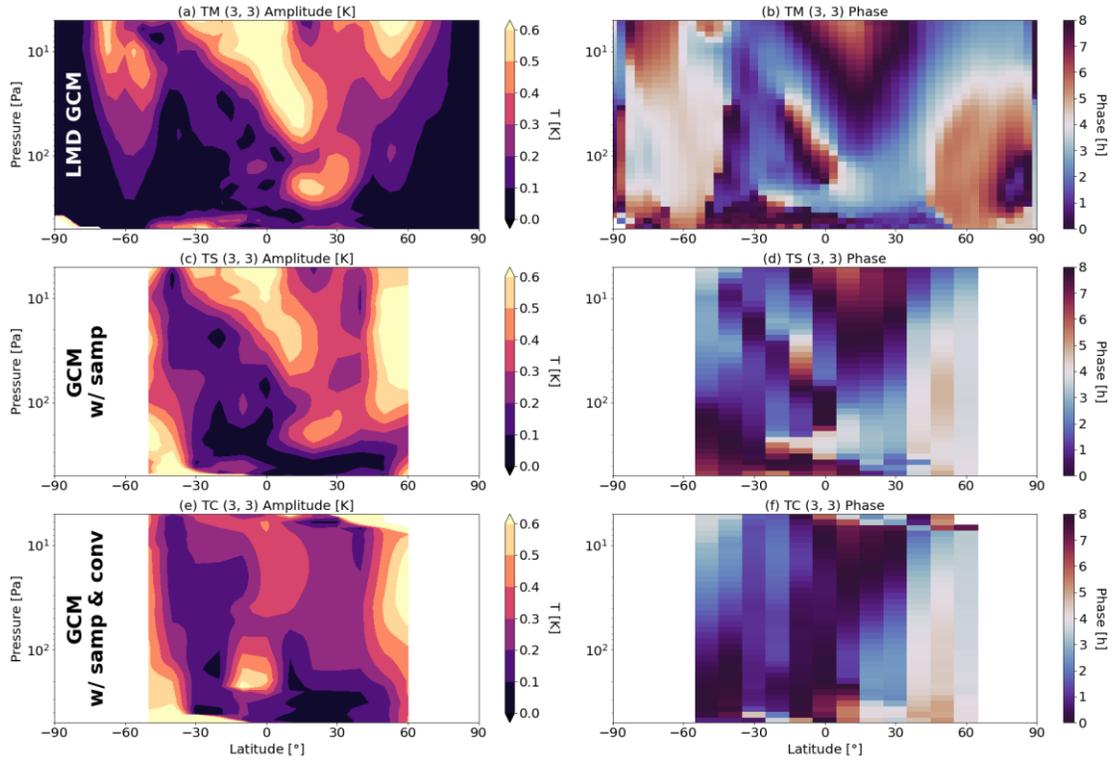

**Figure S9.** Same as Figure S7, but for the ter-diurnal tide.